\documentclass[12pt,draftcls,onecolumn]{IEEEtran}

\def \be {\begin{equation}}
\def \ee {\end{equation}}

\def \nn {\nonumber}

\usepackage{ifpdf}

\usepackage[dvips]{graphicx}
\usepackage{color}

\usepackage[cmex10]{amsmath}
\usepackage{amssymb}

\begin{document}
%
\title{Adaptive Control of Robot Manipulators With Uncertain Kinematics and Dynamics}
%
%
%

\author{Hanlei~Wang 
\thanks{
The author is with the Science and Technology on Space Intelligent Control Laboratory,
Beijing Institute of Control Engineering,
Beijing 100190, China (e-mail: hlwang.bice@gmail.com).}
}
\maketitle

\begin{abstract}
In this paper, we investigate the adaptive control problem for robot manipulators with both the uncertain kinematics and dynamics. We propose two adaptive control schemes to realize the objective of task-space trajectory tracking irrespective of the uncertain kinematics and dynamics. The proposed controllers have the desirable separation property, and we also show that the first adaptive controller with appropriate modifications can yield improved performance, without the expense of conservative gain choice. The performance of the proposed controllers is shown by numerical simulations.
\end{abstract}


%
\IEEEpeerreviewmaketitle

\section{Introduction}

The study on the adaptive control of robot manipulators with dynamic parameter uncertainty has a long and rich history (see, e.g., the early results in \cite{Craig1987_IJRR,Slotine1987_IJRR,Middleton1988_SCL}), and the employment of adaptive control provides robot manipulators with the ability of performing tasks in the unknown environment. The recent advances in adaptive robot control occur in \cite{Cheah2003_TRA,Cheah2006_IJRR,Cheah2006_TAC,Braganza2005_CDC,Dixon2007_TAC} aiming at handling the kinematic parameter uncertainty. Kinematic uncertainty is frequently encountered as the robots perform various work in the task space (e.g., Cartesian space or image space) (see, e.g., \cite{Cheah2003_TRA,Liu2006_TRO}), among which is the now actively studied visual servoing problem (see, e.g., \cite{Liu2006_TRO,Wang2010_TCST,Wang2015_Aut}). These control schemes (e.g., \cite{Cheah2003_TRA,Cheah2006_IJRR,Cheah2006_TAC,Braganza2005_CDC,Dixon2007_TAC,Wang2010_TCST,Wang2015_Aut}) are characterized by the use of an approximate Jacobian matrix (due to the kinematics uncertainty), and the prominent part of the control scheme may be the {approximate transpose Jacobian control} with/without a {kinematic parameter adaptation law}. 

 At the present stage, one may say that the stability properties of the adaptive Jacobian control system under both the uncertain kinematics and dynamics are fully addressed, as can be seen in the above mentioned results, yet it remains unclear about the performance of the system in the sense that some performance issues regarding, e.g., tracking accuracy and transient response, are not adequately studied. In fact, the performance of the now commonly adopted transpose Jacobian feedback (e.g., \cite{Cheah2006_IJRR,Cheah2006_TAC,Braganza2005_CDC,Wang2010_TCST,Wang2015_Aut}), as stated in \cite{Craig2005_Book}, is not desirable especially when the manipulator moves in a large range although the transpose Jacobian feedback for robot task-space control problem shows excellent stability property (refer to the pioneering work in \cite{Takegaki1981_ASME} on the regulation problem and to \cite{Cheah2006_IJRR,Cheah2006_TAC} on the tracking problem). Another commonly adopted task-space control approach is inverse Jacobian feedback (see, e.g., \cite{Craig2005_Book}), and the stability analysis of the inverse Jacobian feedback for regulation problem is given in \cite{Cheah2005_TRO}, which seems much more involved than that of its counterpart (i.e., transpose Jacobian feedback).

 It is well known that the performance of a linear time-invariant system is ensured by appropriately designating the poles of the closed-loop system. For the nonlinear robotic system, this is almost not achievable except for the known parameter case (e.g., the standard computed torque control can result in a linear error dynamics with guaranteed performance---see \cite{Spong2006_Book,Slotine1991_Book}). {Let us now contemplate the standard control problem for a frictionless mass that is governed by $m\ddot{y}=u$, where $y\in R$ denotes the position of the mass, $m\in R$ the mass and $u\in R$ the control input. From the standard linear system theory, if the control $u$ takes a PD action, the design of the gains must take into account the mass property of the system and one advisable design is to choose mass-dependent gains. This standard idea has already appeared in the robot control problem with or without dynamic uncertainties, e.g., the computed torque control actually takes inertia-matrix-dependent PD control plus certain feedforward terms (see, e.g., \cite{Spong2006_Book,Slotine1991_Book}), and the adaptive control in \cite{Slotine1989_Aut} chooses the feedback gain based on the estimated inertia matrix (see \cite[Sec. 3.2]{Slotine1989_Aut}).} However, it is unclear {how to ensure the performance of the robot system under both the kinematic and dynamic uncertainties}. There is also some work addressing the performance in the robust control framework (e.g., \cite{Yao1996_ASME}), yet the gain selection is conservative.


 In this paper, we propose a separation approach to the adaptive control problem for robots with both the uncertain kinematics and dynamics, and two adaptive controllers are proposed. The proposed first controller can also ensure, in the sense of certainty equivalence, the performance of the closed-loop system with essentially the same modification as in \cite{Slotine1989_Aut}. The superior/desirable properties of the proposed approach are summarized as follows.
  \begin{enumerate}
\item It realizes the separation of the kinematic and dynamic loops (i.e., the separation is realized in the case that the joint velocity tracking error is guaranteed to be square-integrable and bounded) thanks to the employment of the kinematic parameter adaptation law (or the new definition of the joint reference velocity) and that of the control law (with the same structure as the Slotine and Li adaptive scheme in the task space \cite[Sec.~3]{Slotine1987_IJRR}) that does not use the approximate transpose Jacobian matrix, while the two loops are coupled and mixed in most existing results (e.g., \cite{Cheah2006_IJRR,Cheah2006_TAC,Braganza2005_CDC,Wang2013_TAC});


    \item the proposed first controller with appropriate modifications that follow the result in \cite{Slotine1989_Aut} improves the performance of the closed-loop system, extending the scheme in \cite{Slotine1989_Aut} to be capable of handling both the kinematic and dynamic uncertainties, {without the expense of conservative gain selection (e.g., \cite{Yao1996_ASME})}.  It is also shown that even under constant-gain feedback, the proposed controller tends to give better performance than the approximate transpose Jacobian feedback.

 \end{enumerate}
We would like to emphasize that the separation property stated in 1) becomes more prominent in industrial robotic applications in that the joint velocity control mode is very common in most industrial manipulators. {Under the joint velocity control mode, we cannot modify the joint servoing module and what we can design is the joint velocity command}. {The separation property of the proposed controllers makes one reduced case of our main result serve well for this application scenario [i.e., taking the joint reference velocity as the joint velocity command of the joint servoing module (see Remark 3)], while the adaptive transpose Jacobian control does not fit this circumstance due to the coupling nature of the adaptive transpose Jacobian feedback in the torque input}.

\section{Kinematics and Dynamics}

Let $x\in R^n$ be the position of the end-effector in the task space (e.g., Cartesian space or image space), and it is relevant to the joint position via the nonlinear mapping \cite{Craig2005_Book,Spong2006_Book}
\be
\label{eq1}
x=f(q)
\ee
where $q\in R^n$ denotes the joint position, and $f: R^n\to R^n$ is the mapping from joint space to task space.

Differentiating (\ref{eq1}) with respect to time gives the relation between the task-space velocity and joint-space velocity \cite{Craig2005_Book,Spong2006_Book}
\be
\label{eq2}
\dot x=J(q)\dot q
\ee
where $J(q)\in R^{n\times n}$ is the Jacobian matrix. In the case that the kinematic parameters are unknown, we cannot obtain the task-space position/velocity by the direct kinematics given above. Instead, we assume that certain task-space sensors (e.g., a camera) are employed to give the task-space position/velocity information. The kinematics (\ref{eq2}) has the following linearity-in-parameters property \cite{Cheah2006_IJRR}.

\emph{Property 1:} The kinematics (\ref{eq2}) depends linearly on a constant parameter vector $a_k$, which gives rise to
\be
\label{eq:a1}
J(q)\xi=Y_k(q,\xi)a_k
\ee
where $\xi\in R^n$ is a vector and $Y_k(q,\xi)$ is the kinematic regressor matrix.

The equations of motion of the manipulator can be written as \cite{Slotine1991_Book,Spong2006_Book}
\begin{equation}
\label{eq3}
M(q)\ddot q+C(q,\dot q)\dot q+g(q)=\tau
\end{equation}
where $M \left( {q }
\right) \in R^{n\times n}$ is the inertia matrix, $C \left( {q ,\dot
{q}} \right) \in R^{n\times n}$ is the Coriolis and centrifugal matrix,
$g \left( {q } \right) \in R^n$ is the gravitational torque, and
$\tau  \in R^n$ is the joint control torque. For the convenience of later reference, three well-understood properties associated with the dynamics (\ref{eq3}) are listed as follows (see, e.g., \cite{Slotine1991_Book,Spong2006_Book}).

\textit{Property 2:} The inertia matrix $M (q )$ is symmetric and uniformly positive
definite.

\textit{Property 3: }The Coriolis and centrifugal matrix $C (q ,\dot q )$ can be
appropriately determined such that $\dot {M}(q) - 2C(q,\dot q)$ is skew-symmetric.

\textit{Property 4: }The dynamics (\ref{eq3}) depends linearly on a
constant parameter vector $a_d $, which leads to
\begin{equation}
\label{eq4}
M \left( q  \right)\dot \zeta  + C \left( q ,\dot {q} \right)\zeta + g \left( {q } \right) = Y_d ( q ,\dot {q}
,\zeta ,\dot \zeta )a_d
\end{equation}
where $\zeta \in R^n$ is a differentiable vector, $\dot {\zeta }$ is the time derivative of $\zeta $, and $Y_d( {q ,\dot {q} ,\zeta ,\dot {\zeta }} )$ is the dynamic
regressor matrix.

\section{Adaptive Control}

In this section, we investigate the adaptive controller design for the robot manipulator given by (\ref{eq2}) and (\ref{eq3}), and the control objective is to drive the robot end-effector to asymptotically track a desired trajectory in the task space, i.e., to ensure that $x-x_d\to 0$ as $t\to\infty$, where $x_d$ denotes the desired task-space trajectory and it is assumed that $x_d$, $\dot x_d$ and $\ddot x_d$ are all bounded.

\subsection{Adaptive Controller I}

Following \cite{Cheah2006_IJRR,Cheah2006_TAC}, we define a joint reference velocity using the estimated Jacobian matrix as
\be
\label{eq6}
\dot q_r =\hat{J}^{-1}(q)\dot x_r
\ee
where $\dot x_r=\dot x_d- \alpha \Delta x$, $\Delta x=x-x_d$ denotes the task-space position tracking error, $\alpha$ is a positive design constant, and $\hat{J}(q)$ is the estimated Jacobian matrix which is obtained by replacing $a_k$ in $J(q)$ with its estimate $\hat{a}_k$.
Differentiating equation (\ref{eq6}) with respect to time gives the joint reference acceleration
\be
\label{eq7}
\ddot q_r=\hat J^{-1}(q)\Big[\ddot x_r-\dot {\hat J}(q)\dot q_r\Big].
\ee
Let us now define a sliding vector
\be
\label{eq8}
s=\dot q-\dot q_r
\ee
and using (\ref{eq2}), (\ref{eq:a1}), and (\ref{eq6}), we can rewrite equation (\ref{eq8}) as
\begin{align}
\label{eq9}
s=&J^{-1}(q)\left[\dot x-J(q)\dot q_r\right]=J^{-1}(q)\left[\dot x-\hat{J}(q)\dot q_r+Y_k(q,\dot q_r)\Delta a_k\right]\nn\\
=& J^{-1}(q)\left[\Delta \dot x+\alpha \Delta x+Y_k(q,\dot{q}_r)\Delta a_k\right]
\end{align}
which can further be written as \cite{Ma1995_ICRA,Cheng2009_Aut}
\be
\label{eq10}
\Delta \dot x=-\alpha\Delta x-Y_k(q,\dot q_r)\Delta a_k+J(q)s
\ee
where $\Delta a_k=\hat{a}_k-a_k$ is the kinematic parameter estimation error.


The control law is given as
\be
\label{eq11}
\tau= -K s+ Y_d(q,\dot q,\dot q_r,\ddot q_r)\hat{a}_d
\ee
where $K\in R^{n\times n}$ is a symmetric positive definite matrix. The estimated dynamic parameter $\hat{a}_d$ (i.e., the estimate of $a_d$) is updated by
\begin{align}
\label{eq12}
\dot{\hat{a}}_d=&-\Gamma_d Y_d^T(q,\dot q,\dot q_r,\ddot q_r)s
\end{align}
where $\Gamma_d$ is a symmetric positive definite matrix. The estimated kinematic parameter $\hat{a}_k$ is updated by the direct adaptation law
\be
\label{eq13}
\dot{\hat{a}}_k=\Gamma_k Y_k^T(q,\dot q_r)[(\beta/\alpha)\Delta \dot x+\Delta x]
\ee
 where $\Gamma_k$ is a symmetric positive definite matrix and $\beta\in[0,1]$ is a design constant.

\emph{Remark 1:} The differences between the adaptive controller here and the one in \cite{Cheah2006_IJRR,Cheah2006_TAC} are that 1) the feedback part in (\ref{eq11}) can be rewritten as $-K J^{-1}(q)\left[\Delta \dot x +\alpha \Delta x+Y_k(q,\dot q_r)\Delta a_k\right]$, which can thus be intuitively interpreted as {inverse Jacobian feedback of both the task-space tracking error and the kinematic parameter estimation error} rather than the approximate transpose Jacobian feedback and 2) the kinematic parameter adaptation law (\ref{eq13}) is based on an adaptive regressor that depends on the joint reference velocity $\dot q_r$ rather than $\dot q$.  The dynamic parameter adaptation law (\ref{eq12}) is actually the same as the one in \cite{Cheah2006_TAC}. The control law (\ref{eq11}) expands the inverse Jacobian based task-space adaptive scheme in \cite[Sec.~3]{Slotine1987_IJRR} to additionally include the inverse Jacobian feedback of the kinematic parameter estimation error, which supplies our controller with the ability of handling the kinematic uncertainties.

Substituting the control law (\ref{eq11}) into the dynamics (\ref{eq3}) yields
\be
\label{eq14}
M(q)\dot s+C(q,\dot q)s=-K s +Y_d(q,\dot q,\dot q_r,\ddot q_r)\Delta a_d
\ee
where $\Delta a_d=\hat{a}_d-a_d$. The closed-loop robotic system can be described by
\be
\label{eq15}
\begin{cases}
\Delta \dot x=-\alpha\Delta x-Y_k(q,\dot q_r)\Delta a_k+J(q)s,
\\
M(q)\dot s+C(q,\dot q)s=-K s +Y_d(q,\dot q,\dot q_r,\ddot q_r)\Delta a_d,
\end{cases}
\ee
and the adaptation laws (\ref{eq12}) and (\ref{eq13}). 

We are presently ready to formulate the following theorem.

\emph{Theorem 1:} {Suppose that the estimated Jacobian matrix $\hat J(q)$ is nonsingular and that all joints of the manipulator are revolute. Then, the control law (\ref{eq11}), the dynamic parameter adaptation law (\ref{eq12}), and the kinematic parameter adaptation law (\ref{eq13}) ensure that the task-space tracking errors converge to zero, i.e., $\Delta x\to 0$ and $\Delta \dot x\to 0$ as $t\to\infty$.}

\emph{Proof: } Following \cite{Slotine1987_IJRR,Ortega1989_Aut}, we take into account the Lyapunov-like function candidate $V_1=(1/2)s^T M(q)s+(1/2)\Delta a_d^T \Gamma_d^{-1}\Delta a_d$, and differentiating $V_1$ with respect to time along the trajectories of the second subsystem in (\ref{eq15}) and of the adaptation law (\ref{eq12}) and using {Property 3}, we obtain
$
\dot V_1=-s^T K s\le 0
$, which implies that $s\in {\cal L}_2\cap {\cal L}_\infty$ and  $\hat a_d\in {\cal L}_\infty$.

Since $J(q)$ is bounded (by the assumption that all joints of the manipulator are revolute), we have that $J(q)s\in {\cal L}_2$, and thus, there exists a constant $l_M>0$ such that $\int_0^t s^T J^T(q)J(q)sdr\le l_M$ for all $t\ge 0$. Then, consider the following quasi-Lyapunov function candidate
\begin{align}
\label{eq17}
V_2=&\frac{1-\beta}{2}\Delta x^T \Delta x+\frac{1}{2\alpha}\bigg[{l_M-\int_0^t s^T J^T(q)J(q)sdr}\bigg]\nn\\
&+\frac{1}{2}\Delta a_k^T \Gamma_k^{-1}\Delta a_k
\end{align}
where the second term of $V_2$ follows the result in \cite[p.~118]{Lozano2000_Book}, and taking the derivative of $V_2$ along the first subsystem in (\ref{eq15}) gives
\begin{align}
\label{eq18}
\dot{V}_2=&-\alpha(1-\beta)\Delta x^T\Delta x- (1-\beta)\Delta a_k^T Y_k^T(q,\dot q_r)\Delta x+\Delta a_k^T \Gamma_k^{-1}\dot{\hat a}_k\nn\\
&+(1-\beta)\Delta x^T J(q)s-\frac{1-\beta}{2\alpha}s^T J^T(q)J(q)s\nn\\
&-\frac{\lambda}{2\alpha} s^T J^T(q)J(q)s.
\end{align}
Using $J(q)s=\Delta \dot x+\alpha\Delta x+Y_k(q,\dot q_r)\Delta a_k$ [from (\ref{eq10})], we can write (\ref{eq18}) as
\begin{align}
\label{eq:a5}
\dot{V}_2=&-\alpha(1-\beta)\Delta x^T\Delta x- (1-\beta)\Delta a_k^T Y_k^T(q,\dot q_r)\Delta x+\Delta a_k^T \Gamma_k^{-1}\dot{\hat a}_k\nn\\
&+(1-\beta)\Delta x^T J(q)s-\frac{1-\beta}{2\alpha}s^T J^T(q)J(q)s\nn\\
&-\frac{\beta}{2\alpha} \big(\Delta \dot x+\alpha\Delta x\big)^T\big(\Delta \dot x+\alpha\Delta x\big)\nn\\
&-\Delta a_k^T Y_k(q,\dot q_r)[(\beta/\alpha)\Delta \dot x+\beta\Delta x]\nn\\
&-\frac{\beta}{2\alpha}\Delta a_k^T Y_k^T(q,\dot q_r)Y_k(q,\dot q_r)\Delta a_k.
\end{align}
Using the inequality
$
\Delta x^T J(q)s\le (\alpha/2)\Delta x^T \Delta x+\left[1/(2\alpha)\right]s^T J^T(q)J(q)s
$ from the standard result concerning the basic inequalities
and substituting the adaptation law (\ref{eq13}) into
equation (\ref{eq:a5}) yields
\be
\label{eq:a6}
\dot{V}_2\le-\frac{\alpha(1-\beta)}{2}\Delta x^T\Delta x-\frac{\beta}{2\alpha} \big(\Delta \dot x+\alpha\Delta x\big)^T\big(\Delta \dot x+\alpha\Delta x\big)\le 0
\ee
which directly gives the result that $\hat{a}_k\in {\cal L}_\infty$ and that $\Delta x\in {\cal L}_2\cap {\cal L}_\infty$ in the case $\beta<1$. In the case $\beta=1$, we obtain from (\ref{eq:a6}) that $\Delta\dot x+\alpha\Delta x\in {\mathcal L}_2$, and further that $\Delta x\in {\mathcal L}_2\cap{\mathcal L}_\infty$ according to the input-output properties of exponentially stable and strictly proper linear systems \cite[p.~59]{Desoer1975_Book}.

From equation (\ref{eq6}), {if the estimated Jacobian matrix $\hat{J}(q)$ is nonsingular}, we have that $\dot{q}_r\in {\cal L}_\infty$ since $\dot{x}_r\in {\cal L}_\infty$. Then, we obtain that  $\dot q\in {\cal L}_\infty$ since $s\in {\cal L}_\infty$, and that $\dot x\in{\cal L}_\infty$ based on (\ref{eq2}). Therefore, $\Delta x$ must be uniformly continuous, and from the properties of square-integrable and uniformly continuous functions \cite[p.~117]{Lozano2000_Book}, we obtain $\Delta x\to 0$ as $t\to\infty$. From (\ref{eq13}), we have that $\dot{\hat{a}}_k\in \cal L_\infty$ since $Y_k(q,\dot q_r)$ and $\Delta x$ are both bounded, which then implies the boundedness of $\dot{\hat{J}}(q)$. Thus, from (\ref{eq7}), we obtain that $\ddot q_r\in {\cal L}_\infty$. From (\ref{eq14}), we obtain that $\dot s\in {\cal L}_\infty$ by using {Property 2}. This leads us to obtain that $\ddot q=\ddot q_r+\dot s\in {\cal L}_\infty$, and that $\ddot x\in {\cal L}_\infty$ from the differentiation of equation (\ref{eq2}), i.e., $\ddot x=J(q)\ddot q+\dot J (q)\dot q$. Therefore, $\Delta \ddot x\in {\cal L}_\infty$, and then $\Delta \dot x$ is uniformly continuous. Due to the result that $\Delta x\to 0$ as $t\to\infty$, we obtain from Barbalat's Lemma \cite{Slotine1991_Book} that $\Delta \dot x\to 0$ as $t\to\infty$. \hfill {\small $\blacksquare$}

\subsection{Adaptive Controller II}

We now present an adaptive controller that also has the separation property but uses different joint reference velocity and kinematic parameter adaptation law. This controller relies on the joint reference velocity defined as
\be
\label{eq:a8}
\dot q_r=\hat J^{-1}(q)\dot x_d-\alpha \hat J^T(q)\Delta x.
\ee
We then have that
\be
\label{eq:a9}
\Delta \dot x=-\alpha\hat J(q)\hat J^T (q)\Delta x-Y_k(q,\dot q)\Delta a_k+\hat J(q)s.
\ee

The control law and the dynamic parameter adaptation law are still (\ref{eq11}) and (\ref{eq12}) yet with $\dot q_r$ given by (\ref{eq:a8}). The kinematic parameter adaptation law is given as
\be
\label{eq:a10}
\dot{\hat a}_k=\Gamma_k Y_k^T(q,\dot q)\Delta x.
\ee

\emph{Theorem 2:} The control law (\ref{eq11}) and dynamic parameter adaptation law (\ref{eq12}) with $\dot q_r$ given by (\ref{eq:a8}), and kinematic parameter adaptation law (\ref{eq:a10}) ensure that  $\Delta x\to 0$ and $\Delta \dot x\to0$ as $t\to\infty$ provided that the estimated Jacobian matrix $\hat J(q)$ is nonsingular.

\emph{Proof:} For the dynamic loop, we can directly obtain that $s\in{\mathcal L}_2\cap {\mathcal L}_\infty$ and $\hat a_d\in{\mathcal L}_\infty$, by following similar procedures as in the proof of Theorem 1. Then, there exists a positive constant $l_M^\ast$ such that $\int_0^t s^T(r)s(r)dr\le l_M^\ast$, $\forall t\ge 0$. Consider the following quasi-Lyapunov function candidate
\be
\label{eq:a11}
V_2^\ast=\frac{1}{2}\Delta x^T \Delta x+\frac{1}{2\alpha}\left[l_M^\ast-\int_0^t s^T(r)s(r)dr\right]+\frac{1}{2}\Delta a_k^T\Gamma_k^{-1}\Delta a_k
\ee
and we obtain
\be
\dot V_2^\ast\le -(\alpha/2)\Delta x^T \hat J(q)\hat J^T(q)\Delta x\le 0
\ee
where we have used the following result from the standard basic inequalities
\be
\Delta x^T\hat J(q)s\le (\alpha/2)\Delta x^T \hat J(q)\hat J^T(q)\Delta x+1/(2\alpha)s^T s.
\ee
Then, we can show the convergence of the task-space tracking errors, using similar procedures as in the proof of Theorem 1. \hfill {\small $\blacksquare$}

\emph{Remark 2:} The second adaptive controller achieves separation by using a strong feedback of the tracking error $\Delta x$ in the definition of the joint reference velocity (this form of reference velocity appears in the context of global task-space control with known kinematic parameters \cite{Li2013_Aut}). This can be noticed more clearly from (\ref{eq:a9}) and the final equivalent feedback gain is $\alpha\hat J(q)\hat J^T(q)$ at the task-space velocity level. Due to this, the second adaptive controller is applicable to robots with prismatic joints, in contrast with the proposed first adaptive controller. The kinematic parameter adaptation law (\ref{eq:a10}) is the same as the direct version of that in \cite{Cheah2006_IJRR} (i.e., with the prediction error being removed), and the separation of the kinematic and dynamic loops is mainly attributed to the new definition of the joint reference velocity (\ref{eq:a8}) and the control law (\ref{eq11}).

\emph{Remark 3:} From the separation analysis in the proofs of Theorem 1 and Theorem 2, we can obtain two adaptive kinematic schemes. One is given by the joint reference velocity (\ref{eq6}) and kinematic parameter adaptation law (\ref{eq13}), and the other is given by  (\ref{eq:a8}) and (\ref{eq:a10}). Both of them are expected to serve well for industrial robotic applications. In fact, the joint velocity servoing module can generally ensure that the joint velocity tends sufficiently fast to the joint reference velocity in the sense that the joint velocity tracking error $s=\dot q-\dot q_r$ is square-integrable and bounded. Hence, $J(q)s\in {\mathcal L}_2\cap {\mathcal L}_\infty$ and $s\in{\mathcal L}_2\cap {\mathcal L}_\infty$. Consider the same quasi-Lyapunov function as (\ref{eq17}) or (\ref{eq:a11}), and then the tracking error convergence can be obtained by following similar analysis as in the proof of Theorem 1 or Theorem 2.

\emph{Remark 4:} {The assumption that the manipulator is away from the singular configuration and the use of the parameter projection algorithms ensure that the estimated Jacobian matrix $\hat J(q)$ is nonsingular in the parameter adaptation process (see, e.g., \cite{Cheah2006_IJRR,Cheah2006_TAC,Dixon2007_TAC})}.

\section{Performance of the System}

 In this section, we show how the first adaptive controller given in Sec. III improves the performance of the robotic system under both the uncertain kinematics and dynamics by a suitable nonconservative modification. This modification follows \cite{Slotine1989_Aut}, yet in the context of task-space robot control with kinematic uncertainties. The extension turns out to be direct thanks to the formulation in the previous section, yet, here, our emphasis is on demonstrating {why this modification implies potentially good performance}. Moreover, it will be shown that even under constant-gain feedback, inverse Jacobian feedback yields potentially better performance than transpose Jacobian feedback.

Following \cite{Slotine1989_Aut}, we specify the feedback gain $K$ in (\ref{eq11}) as
\be
\label{eq21}
K=\lambda_c \hat M(q)
\ee
with $\lambda_c$ being a positive design constant and meanwhile modify the adaptation law (\ref{eq12}) as
\be
\label{eq22}
\dot{\hat a}_d=-\Gamma_d Y_d^T(q,\dot q,\dot q_r,\ddot q_r^\ast)s
\ee
  where $\hat M(q)$ is the estimated inertia matrix obtained by replacing $a_d$ in $M(q)$ with $\hat a_d$, and $\ddot q_r^\ast=\ddot q_r-\lambda_c s$. The selection (\ref{eq21}) and the modification (\ref{eq22}) yields (the same as the case in \cite{Slotine1989_Aut})
$
\dot V_1=-\lambda_c s^T M(q)s\le 0.
$
Then, the same result as in {Theorem 1} follows. 

{Let us now focus on interpreting the performance issues from two perspectives. First, the derivative of $V_1$ can further be written as
\begin{align}\dot V_1
=-\lambda_C\underbrace{\psi^T J^{-T}(q)M(q) J^{-1}(q)\psi}_{V^\ast}
\end{align}
with $\psi=\Delta \dot x +\alpha \Delta x+Y_k(q,\dot q_r)\Delta a_k$, which implies the exponential convergence of $\Delta \dot x+\alpha\Delta x+Y_k(q,\dot q_r)\Delta a_k$ with the rate $\lambda_c$ in the case that the dynamic parameter is known [in this case, $V_1=(1/2)V^\ast$] and further the exponential convergence of $\Delta x$ with the rate $\min\left\{\lambda_c,\alpha\right\}$ in the case that the kinematic parameter is known (i.e., the performance is quantified in the sense of certainty equivalence).}
On the other hand, based on (\ref{eq9}), we can rewrite the definitions of $\dot q_r,\ddot q_r$ in (\ref{eq6}) and (\ref{eq7}) as
\begin{align}
\dot q_r=&J^{-1}(q)\left[\dot x_r-Y_k(q,\dot q_r)\Delta a_k\right]
\\
\ddot q_r=&J^{-1}(q)\left[\ddot x_r-\dot Y_k(q,\dot q_r)\Delta a_k-Y_k(q,\dot{q}_r)\dot{\hat{a}}_k\right]-J^{-1}(q)\dot J(q)\dot q_r.
\end{align}
With the above new formulation of $\dot q_r$ and $\ddot q_r$, the control law can now be written as
\begin{align}
\label{eq26}
\tau=& \hat M(q) J^{-1}(q)\Big[\ddot x_d-(\alpha+\lambda_c)\Delta \dot x-\alpha\lambda_C\Delta x\nn\\
&-\left(\dot{Y}_k(q,\dot q_r)+\lambda_c Y_k(q,\dot q_r)\right)\Delta a_k-Y_k(q,\dot q_r)\dot{\hat{a}}_k\Big]\nn\\
&+\left[\hat C(q,\dot q)J^{-1}(q)-\hat M(q)J^{-1}(q)\dot J(q)J^{-1}(q)\right]\nn\\
&\times\left[\dot x_r-Y_k(q,\dot q_r)\Delta a_k\right]+\hat g(q).
\end{align}
{which is quite similar to the certainty-equivalence form of the task-space inverse dynamics (see, e.g., \cite{Luh1980_TAC,Craig2005_Book,Spong2006_Book}) and thus ensures the performance in the sense of certainty equivalence, where $\alpha$ and $\lambda_c$ act as the quantification of the performance (e.g., speed of response and robustness)}. We now present the following theorem to summarize the above result.

{\emph{Theorem 3:} The adaptive controller (\ref{eq26}), (\ref{eq22}), and (\ref{eq13}) ensures the exponential convergence of the task-space position tracking error with the rate $\min\{\lambda_c,\alpha\}$ in the sense of certainty equivalence.}

{
To further clarify the potential benefit of the choice of the feedback gain (\ref{eq21}), let us consider a scenario that a convergence rate of the task-space tracking error $\gamma^\ast$ (in the sense of certainty equivalence) is required by certain task. To serve this requirement, we only need to specify $\lambda_c$ and $\alpha$ as $\lambda_c=\alpha=\gamma^\ast$ as using the control scheme here. In contrast, if still using the constant gain feedback (similar to the case in \cite{Yao1996_ASME}), the convergence rate of $s$ becomes $\lambda_{\min}\{K\}/\lambda_{\max}\{M(q)\}$ and thus the best choice is perhaps to specify $K$ and $\alpha$ such that $\alpha=\gamma^\ast$ and $\lambda_{\min}\{K\}\ge\gamma^\ast \lambda_{\max}\{M(q)\}$. Due to the uncertainty of $M(q)$, conservativeness is generally inevitable.}

\emph{Remark 5:} The adaptive controller given by (\ref{eq26}), (\ref{eq22}), and (\ref{eq13}) yields
\be
\label{eq28}
\begin{cases}
\Delta \dot x=-\alpha\Delta x-Y_k(q,\dot q_r)\Delta a_k+J(q)s,\\
M(q)\dot s+C(q,\dot q)s\\
=-\lambda_c M(q) s+Y_d(q,\dot q,\dot q_r,\ddot q_r^\ast)\Delta a_d.
\end{cases}
\ee
The feedback gains in both the systems of (\ref{eq28}) are inertia-dependent (the apparent inertia of the first subsystem in (\ref{eq28}) can be considered as $I_n$). This gives the additional demonstration on why the adaptive scheme given by (\ref{eq26}), (\ref{eq22}), and (\ref{eq13}) implies good task-space tracking performance.
The approximate transpose Jacobian feedback adopted in \cite{Cheah2006_IJRR} [i.e., $-\hat J^T(q) K\hat J(q)s$] with the gain selection (\ref{eq21}) and appropriate modification of the dynamic parameter adaptation law would render the feedback gain (with respect to $s$) in the closed-loop dynamics as $-\lambda_c\hat J^T(q)M(q) \hat J(q)$, which, in most cases, cannot match $M(q)$. This also suggests that for the task-space tracking problem, approximate transpose Jacobian control (e.g., \cite{Cheah2006_IJRR,Cheah2006_TAC}) may not be preferred.

\emph{Remark 6:} The assertion in Remark 5 holds even for the case of constant-gain feedback (i.e., $K$ is chosen to be constant). It is well known that the task-space inertia $J^{-T}(q)M(q)J^{-1}(q)$ involves the inversion of the Jacobian matrix. In the case of using approximate transpose Jacobian feedback as is the case in \cite{Cheah2006_IJRR,Cheah2006_TAC}, the inversion of the transpose of the approximate Jacobian matrix would cancel the transpose of the approximate Jacobian matrix and render the feedback gain to be $K$, which implies that we have to rely on the constant gain $K$ to compensate for the task-space inertia $J^{-T}(q)M(q)J^{-1}(q)$. In the case of using inverse Jacobian feedback without involving the transpose of the approximate Jacobian matrix as in our result, the task-space formulation renders the inverse Jacobian feedback premultiplied by $J^{-T}(q)$, i.e., using the Jacobian-dependent varying gain $J^{-T}(q)K J^{-1}(q)$ to compensate for the varying task-space inertia $J^{-T}(q)M(q)J^{-1}(q)$, which tends to be much easier. The performance superiority of inverse Jacobian feedback control is thus obvious.

\emph{Remark 7:}  In the visual tracking problem for robots with uncertainties in the camera model and/or manipulator kinematics, most results, e.g., \cite{Cheah2007_ICRA,Wang2010_TCST,Wang2012b_Mech,Li2012_ACC,Wang2015_Aut}, are fully/partly based on the approximate transpose Jacobian feedback. The use of constant-gain feedback in the joint space (i.e., in the form $-K s$) occurs \cite{Lizarralde2013_Aut} [which employs the indirect kinematic parameter adaptation law, and additionally require the persistent excitation of the kinematic regressor $Y_k(q,\dot q)$ so that the convergence of the tracking error is ensured], and also appears in \cite{Braganza2005_CDC,Wang2010_TCST,Wang2012b_Mech,Li2012_ACC} as part of the overall feedback action (the use of $-K s$ alone in this case, yet, cannot ensure stability), which is the same as the one proposed in our result and may also be interpreted as inverse Jacobian feedback, yet the rationality/interpretation of doing so and the performance issues associated with the closed-loop system are not adequately addressed. The adaptive kinematic regressor used in the first adaptive controller is the same as the work in \cite{Ma1995_ICRA,Cheng2009_Aut} (where the work in \cite{Ma1995_ICRA} handles the control of attitude-controlled space manipulators using adaptive Jacobian technique with a coupled stability analysis for the kinematic and dynamic loops), and furthermore if we removed the velocity tracking error $\Delta \dot x$ in (\ref{eq13}), the kinematic parameter adaptation law (\ref{eq13}) would be the same as the one in \cite{Ma1995_ICRA,Cheng2009_Aut}. {It is worth remarking that the adaptive controller given by (\ref{eq11}), (\ref{eq12}), and (\ref{eq13}) with $\lambda=0$ is quite similar to the one in \cite{Ma1995_ICRA} (with its journal version in \cite{Ma1996_CTA}, which is mainly pursued in the Chinese control literature), and due to the reason of language, it passed out of the knowledge of the international community. The main novel points of our first result given in Sec. III-A, in comparison with \cite{Ma1995_ICRA}, lies in the proposed more general direct kinematic parameter adaptation law, the separation analysis (by using a quasi-Lyapunov analysis), and the clarification of the separation property (rationality) of using inverse Jacobian feedback as well as the adaptive kinematic regressor matrix (which then implies its potential applications to robots having an unmodifiable joint servoing controller yet admitting the design of the joint velocity command).} In addition, the first adaptive controller is demonstrated to be convenient for accommodating the performance issues.




\section{Simulation Results}

Let us consider a standard 2-DOF (degree-of-freedom) planar manipulator that grasps an unknown tool. The physical parameters of the 2-DOF manipulator are not listed for saving space. 
The sampling period is chosen as 5 ms. The desired trajectory of the manipulator end-effector is chosen as
$
x_d=[1.6754+0.3\cos \pi t,
3.9950+0.3\sin\pi t]^T
$.

For the first adaptive controller given in Sec. III-A, the controller parameters $K$, $\alpha$, $\Gamma_d$, and $\Gamma_k$ are chosen as $K=30 I_2$, $\alpha=10$, $\beta=0.5$, $\Gamma_d=200 I_4$, and $\Gamma_k=300 I_3$, respectively. The initial parameter estimates are chosen as $\hat{a}_d(0)=\left[0,0,0,0\right]^T$ and $\hat a_k(0)=\left[4.0,5.0,2.0\right]^T$, while their actual values are $a_d=[7.9628, -0.9600, 19.2828,$
$10.1495]^T$ and $a_k=\left[2.0000,3.3856,0.8000\right]^T$. Simulation results are shown in Fig. 1 and Fig. 2. For the second adaptive controller, the controller parameters are chosen to be the same as the first except that the design parameter $\alpha$ is decreased to $\alpha=1.5$ (since in this case, the equivalent feedback gain contains the transpose of the estimated Jacobian matrix). Simulation results are plotted in Fig. 3 and Fig. 4.

Under the same context, we also conduct the simulation when the controller given in \cite{Cheah2006_IJRR,Cheah2006_TAC} is adopted. The control law in this case employs the approximate transpose Jacobian feedback $-\hat{J}^T(q)K \hat{J}(q)s$ and the kinematic parameter adaptation law takes the form $\dot{\hat{a}}_k=\Gamma_k Y_k^T(q,\dot q)[(\beta/\alpha)\Delta \dot x+\Delta x]$. The controller parameters are chosen to be the same as in the first adaptive controller. Simulation results in this context are shown in Fig. 5 and Fig. 6.

One obvious difference between the simulation results under the first controller and those under the one in \cite{Cheah2006_IJRR,Cheah2006_TAC} is that the first controller results in better tracking accuracy [approximately $0.0015$ m (Fig. 1) versus $0.006$ m (Fig. 5) after $t=6$ s] and more adequate utilization of the joint torques (see Fig. 2 as compared with Fig. 6). The tracking accuracy under the second adaptive controller (after $t=6$ s) is comparable to that under the one in \cite{Cheah2006_IJRR,Cheah2006_TAC}.

The tracking error with the controller under the estimated-inertia-based feedback action $-\lambda_c\hat{M}(q)s$ is shown in Fig. 7, where we choose $\lambda_c=\alpha=10$ so that the closed-loop dynamics is approximate to a critically damped linear dynamics, and the other parameters are chosen to be the same as those in the first adaptive controller. The main superiority may lie in the fact that responses of the tracking errors become more uniform and the tracking errors converge faster as compared with the first adaptive controller using constant-gain feedback (see Fig. 1).

\begin{figure}[!t]
\centering
\begin{minipage}[t]{1\linewidth}
\centering
\includegraphics[width=3.3in]{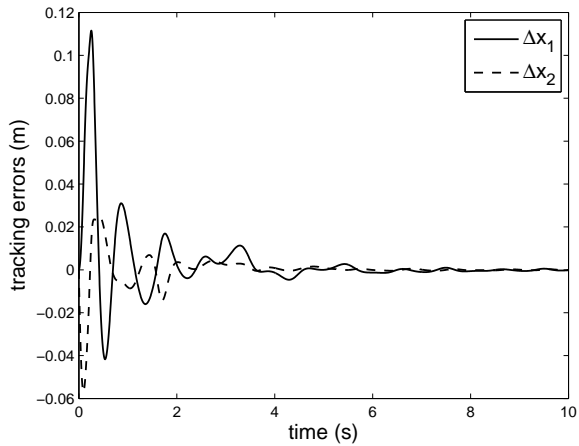}
\caption{End-effector position tracking errors (first adaptive controller).}\label{fig:side:a}
\end{minipage}%
\end{figure}

\begin{figure}[!t]
\centering
\begin{minipage}[t]{1\linewidth}
\centering
\includegraphics[width=3.3in]{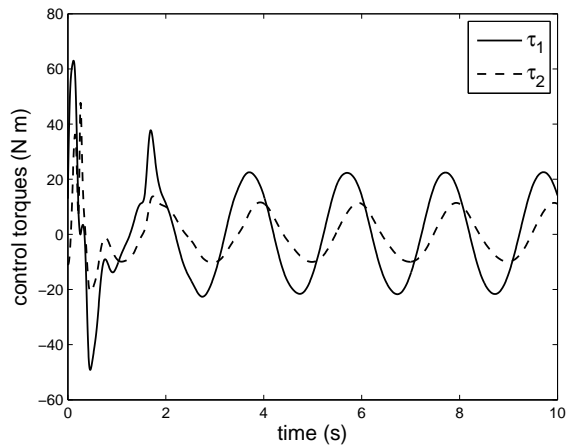}
\caption{Joint control torques (first adaptive controller).} \label{fig:side:b}
\end{minipage}
\end{figure}

\begin{figure}[!t]
\centering
\begin{minipage}[t]{1\linewidth}
\centering
\includegraphics[width=3.3in]{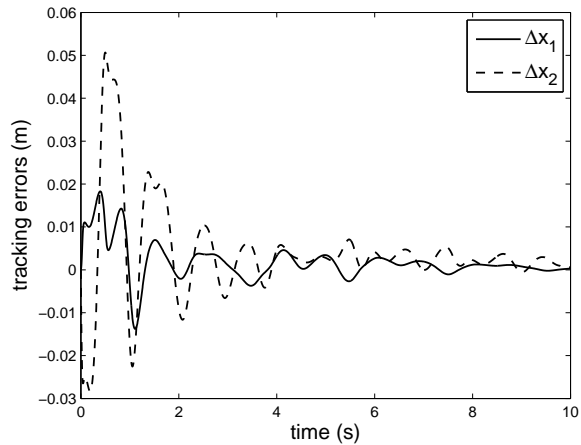}
\caption{End-effector position tracking errors (second adaptive controller).}\label{fig:side:a}
\end{minipage}%
\end{figure}

\begin{figure}[!t]
\centering
\begin{minipage}[t]{1\linewidth}
\centering
\includegraphics[width=3.3in]{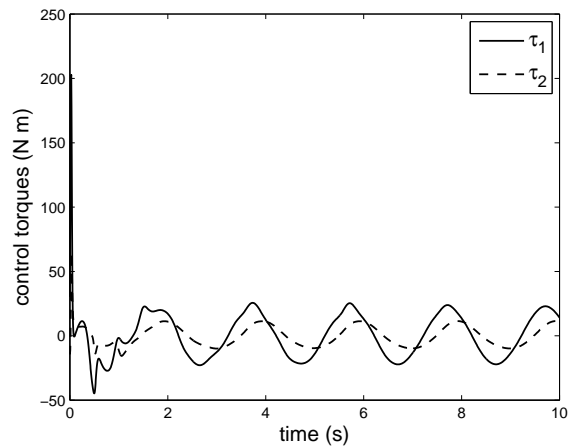}
\caption{Joint control torques (second adaptive controller).} \label{fig:side:b}
\end{minipage}
\end{figure}

\begin{figure}[!t]
\centering
\begin{minipage}[t]{1\linewidth}
\centering
\includegraphics[width=3.3in]{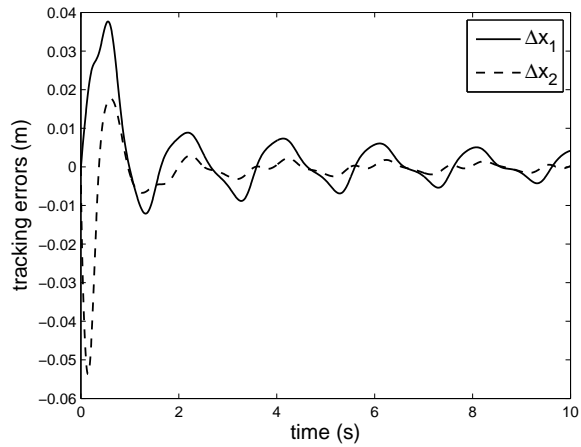}
\caption{End-effector position tracking errors (adaptive transpose Jacobian feedback).}\label{fig:side:a}
\end{minipage}%
\end{figure}

\begin{figure}[!t]
\centering
\begin{minipage}[t]{1\linewidth}
\centering
\includegraphics[width=3.3in]{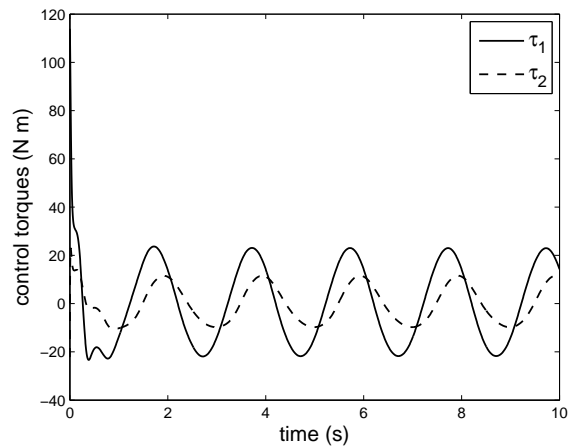}
\caption{Joint control torques (adaptive transpose Jacobian feedback).} \label{fig:side:b}
\end{minipage}
\end{figure}

\begin{figure}[!t]
\centering
\begin{minipage}[t]{1\linewidth}
\centering
\includegraphics[width=3.3in]{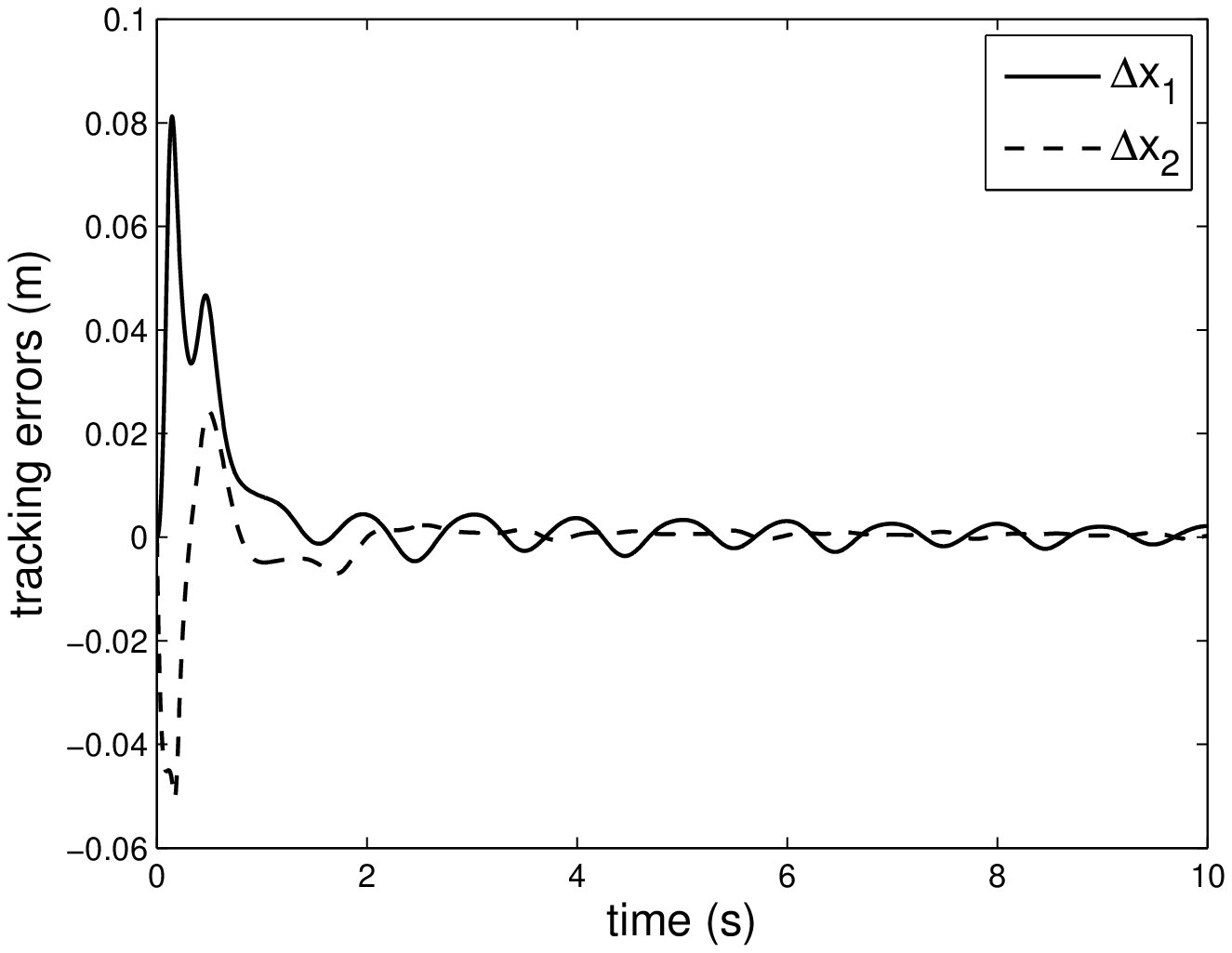}
\caption{End-effector position tracking errors (estimated-inertia-based feedback).} \label{fig:side:b}
\end{minipage}
\end{figure}

\section{Conclusion and Discussion}

In this paper, we consider the adaptive tracking problem for robot manipulators subjected to both the kinematic and dynamic uncertainties. We propose two adaptive controllers that enjoy the separation property. The performance is then shown to be conveniently ensured under the first adaptive controller in the sense of certainty equivalence, with essentially the same modification of the control law and of the dynamic parameter adaptation law as in \cite{Slotine1989_Aut}. Our study also suggests that to obtain a potentially good task-space tracking performance, adaptive inverse Jacobian feedback seems preferable than the commonly adopted adaptive transpose Jacobian feedback (e.g., \cite{Cheah2006_IJRR,Cheah2006_TAC}). 

One desirable feature of the proposed control schemes is that the separation of the kinematic and dynamic loops makes one reduced version of our control scheme rather suitable for industrial robotic applications. {This originates from the fact that the kinematic control law (represented by the joint reference velocity) plus the kinematic parameter adaptation law will ensure the convergence of the task-space tracking errors so long as the joint servoing loop (commonly embedded in most industrial robots) can ensure that the joint velocity tends sufficiently fast to the joint reference velocity in the sense that the joint velocity tracking error is square-integrable and bounded}.


%








\bibliographystyle{IEEEtran}
\bibliography{..//Reference_list_Wang}
%
%
%

%








\end{document}